\documentstyle[aps,preprint,epsfig]{revtex}
\newcommand{\str}{\rule{0ex}{2.7ex}}
\newcommand{\strr}{\rule{0ex}{3.50ex}}

\newcommand{\PP}{$A_{\pi\pi}$}
\newcommand{\PH}{$A_{\pi p}$}
\begin{document}
\title{Inversion Potentials for Meson--Nucleon and Meson--Meson
Interactions%
\footnote{Contribution to the International Conference on Inverse and 
Algebraic Scattering Theory, Lake Balaton '96}}

\author{M. Sander and H.V. von Geramb}
\address{Theoretische Kernphysik, Universit\"at Hamburg,\\
Luruper Chaussee 149, D--22761 Hamburg, Germany}

\maketitle

\begin{abstract}
Two--body interactions of elementary particles are useful
in particle and nuclear physics to describe qualitatively and
quantitatively few-- and many--body systems.
We are extending for this purpose the quantum inversion approach for systems
consisting of nucleons and mesons.
From the wide range of experimentally studied two--body systems we concentrate
here on $\pi N$, $\pi\pi$, $K^+N$, $K \pi$ and $K\bar{K}$.
As input we require results of phase shift analyses.
Quantum inversion Gelfand--Levitan and Marchenko
single and coupled channel algorithms are used for Schr\"odinger type
wave equations in partial wave decomposition.
The motivation of this study
comes from our two approaches: to
generate and investigate potentials directly from data by
means of inversion and alternatively use linear and nonlinear
boson exchange models.
The interesting results of inversion are coordinate space informations
about radial ranges, strengths, long distance behaviors,
resonance characteristics, threshold effects, scattering lengths
and bound state properties.
\end{abstract}
\section{Introduction}
It is a paradigm of particle and nuclear physics to describe complex 
many--body systems in terms of two--body interactions or two--body 
t--matrices. The formulation of a two--body interaction is therefore a 
central goal for generations of physicists, both of the experimental and 
theoretical community. Roughly speaking, we separate these efforts of finding 
two--body interactions into groups which orient themselves very closely on
data, whereas the other extreme 
follows a fundamental approach. In another contribution to
this conference we have dwelled upon the need to follow both
approaches \cite{jae96}.

Encouraged by the tremendous success of the inverse scattering method 
for fixed angular momentum in application to 
nucleon--nucleon interactions 
we extend our study in this contribution to  meson--nucleon
and meson--meson systems in the framework of the Gelfand--Levitan and Marchenko
theory. A comprehensive description of our mathematical and numerical framework
can be found elsewhere [2,3] 
and we shall concentrate here solely on the presentation and interpretation of 
results which are a small fraction of all results contained in the thesis 
\cite{san96}.
In particular we study $\pi N$, $\pi\pi$, $K^+N$, $K \pi$ and $K\bar{K}$
scattering.
As experimental input for the inversion algorithms we used phase shift analyses
of Arndt et al. for $\pi N$ \cite{arnpin} and $K^+N$ \cite{arnkn},
of Froggatt et al. for $\pi\pi$ \cite{fro77} and of
Estabrooks et al. for $K\pi$ \cite{est78}. 
We limit ourselves to subinelastic and subreaction threshold data.
An effective range
parameterization is used for the $K\bar{K}$ system \cite{kam94}.
A critical and comprehensive assessment of the data can be found in 
\cite{san96}.

For the two--body system the relativistic Schr\"odinger equation
\begin{equation} \label{wellengl}
- f_\ell^{\prime\prime} (k,r) + \left( { \ell ( \ell + 1) \over r^2} 
+ 2 \mu(s)  V_\ell(r) \right) f_\ell(k,r) 
= k^2 f_\ell(k,r)
\end{equation}
is assumed to be valid.
The CM parameters are given by
\begin{equation}
k^2 = { \displaystyle m_2^2 ( T_{Lab}^2 + 2 m_1 T_{Lab}) \over
\displaystyle ( m_1 + m_2 )^2 + 2 m_2 T_{Lab}}.
\end{equation}
or
\begin{equation}
k^2 = \left( { s - m_1^2 - m_2^2 \over 2 \sqrt{s}} \right)^2 -
{m_1^2m_2^2 \over s}
\end{equation}
with
\begin{equation}
s = M_{12}^2 = \left( \sqrt{k^2 + m_1^2} + \sqrt{k^2 + m_2^2} \right)^2 
= ( m_1 + m_2 )^2 + 2 m_2 T_{Lab},
\end{equation}
$m_1, m_2$ are the masses of the projectile and the target and 
$T_{Lab}$ is the kinetic energy in the laboratory.
The experimental data are given either as a function of $T_{Lab}$ or the 
Mandelstam variable $s$.
As reduced mass $\mu(s)$ we use either the non--relativistic reduced mass
\begin{equation}
\mu = { m_1 m_2 \over m_1 + m_2 }
\end{equation}
or the reduced energies \cite{bak86}
\begin{equation} 
\mu (s) = \frac12 {dk^2(s) \over d\sqrt{s}} = 
\frac12 { s^2 - (m_1^2 - m_2^2 )^2 \over
2s^{3/2}}.
\end{equation}

This expression (6) is generally approximated by the
low--energy limit or the conventional non--relativistic reduced mass.
Any of these options is to be motivated by the application of the interaction
potentials in other contexts.
Depending on the choice of the reduced mass, we obtain different 
inversion potentials.
The potential is in any case local and energy independent but dependent 
on the channel quantum numbers ($\ell S J T)$. 
Our numerical algorithm guarantees that insertion of the potential 
in eqn.\ (\ref{wellengl})
reproduces the input phase shifts $\delta (k)$ better than 0.02$^\circ$.

The simplicity of our potential operator may be surprising in view of the
non--locality implied by results from meson exchange models.  
It is also our opinion that the actual potential should be non--local,
but we understand that the inversion potential represents a local
equivalent yielding the same on--shell two--body t--matrix 
for the full operator. 
The local potential permits to compute off--shell t--matrices with a 
Lippmann--Schwinger equation with the implication that a non--local
potential may yield a different off--shell continuation than a local potential.
This difference in the off--shell domain can become important in few-- and 
many--body systems or in the interaction regions of two--body wave--functions.
For nucleon--nucleon systems we have put much effort in attempts to clarify
this point and find no evidence for differences in
observables [1,4].   
This is a very surprising result and it is the purpose of this study to
initiate a comparison of strictly equivalent local on--shell potentials 
with their non--local counterparts in the realm of meson--nucleon and 
meson--meson interactions.

\section{$\pi N$ Scattering}
Partial wave phase shifts of the $\pi N$ system are determined
by an analysis of elastic and charge exchange scattering
$\pi^+ p \to \pi^+ p ,
\pi^- p \to  \pi^- p$ and 
$\pi^- p \to \pi^0 n$.
They form a complete set of observables
entering a partial wave analysis with 
isospin $T=\frac12$ and $\frac32$ \cite{arnpin}. The notation 
for the isospin and angular momentum channels is
$ \ell_{2T,2S}$.
We used  the SM95 analysis of Arndt et al. for
channels $\ell \leq 3$  [4,5].

\subsection{$\pi N$ p--Wave Resonances}

The most prominent resonances are the $\Delta(1232)$ in the 
$P_{33}$ and the $N(1440)$ Roper resonance in the $P_{11}$ channels.
The phase shifts for these resonances are shown in the left part of 
Fig. \ref{abb_pin1}. 
We factorize the S--matrix into
a {\em resonant} and a {\em non--resonant} part
$
S (k) = S_r (k) S_p (k).
$
For the resonant part we use 
a resonance and an auxiliary pole parameterization \cite{kuk89}
\begin{equation}
S_r (k) =
{ ( k + k_r ) ( k - k_r^* ) \over (k - k_r ) ( k + k_r^* ) }
{ ( k + k_h ) ( k - k_h^* ) \over (k - k_h ) ( k + k_h^* ) },
\end{equation}
which contains the right amount of zeros and poles for a 
decomposition into Jost functions. The rest $S_p (k)$ of the S--matrix is 
parameterized in our usual parameterization scheme for $\delta_p (k)$
in connection with the symmetric Pad\'e approximant for the exponential
function [3].

In Table \ref{tab_pin2} are summarized the relevant parameters for 
this decomposition using SM95 \cite{arnpin}. 
As shown in Fig. \ref{abb_pin1} (left) this reproduces the input 
phase shifts very well for $T_\pi \leq 500$ MeV.
In Fig. \ref{abb_pin1} (right) we show the inversion potential for the 
$P_{33}$ channel with a repulsive well, enabling  
tunneling  for the pion--nucleon system, with a very strong short range 
attraction,
which yields the $\Delta(1232)$ quasi bound state. 
The long range part of this potential, not visible in Fig. \ref{abb_pin1},
behaves like a Yukawa tail with a strength $Y=650.0$ [MeVfm] 
and $\mu = 1.77$ [fm$^{-1}$], see also section V.
 The relative distance of the
centers of mass in the strong attractive region is unexpectedly small,
$r \leq 0.25$ fm.
In view of the large radii of the charge form factor of the pion and the 
nucleon, approximately 0.54 fm for the pion and 
0.7 fm for the nucleon, this relative 
distance implies more than 90\% overlap of the intrinsic structures 
before the strong attractive potential simulates a phase transition of the
pion--nucleon quark content into the 3--quark content of the $\Delta$.
Ultimately such explanation must be confirmed by QCD calculations.
We decline and warn from a far reaching interpretation of this 
potential with its strength and radial dimensions. 

\subsection{$\pi N$ Scattering Lengths and $\pi NN$ Coupling Constants}

From the $\pi N$ potentials in the $T= \frac12,\frac32$ s--channels we find
the scattering lengths
$a_1   =  0.178 \, [m_\pi^{-1}],$ $ a_3 = -0.088 \, [m_\pi^{-1}]$.
For a comparison with several other predictions see Table \ref{tab_pins}.
These results may be used in the Goldberger--Miyazawa--Oehme
sum rule \cite{GMO}
\begin{equation}
{\displaystyle f_{\pi NN}^2 \over \displaystyle 4 \pi} =
{\displaystyle  (m_\pi^2 - \mu^2) ( m_N + m_\pi) \over
\displaystyle 6 m_N  m_\pi}
( a_1 - a_3) -
{\displaystyle  m_\pi^2 - \mu^2 \over \displaystyle 8 \pi^2}
\int_0^\infty {\displaystyle \sigma_{\pi^-p} - \sigma_{\pi^+p}
\over \sqrt{ \displaystyle q^2 + m_\pi^2}} dq
\end{equation}
to obtain a model independent estimate for the $\pi NN$ coupling constant.
Using the simplified form of this sum rule \cite{wor92}
\begin{equation}
{\displaystyle f_{\pi NN}^2 \over \displaystyle 4 \pi} =
 0.19 m_\pi ( a_1 - a_3) - (0.025 \mbox{ mb}^{-1})
{\cal J},
\end{equation}
where the integral
\begin{equation}
{\cal J} = {\displaystyle  1 \over \displaystyle 4 \pi^2}
\int_0^\infty {\displaystyle \sigma_{\pi^-p} - \sigma_{\pi^+p}
\over \sqrt{\displaystyle q^2 + m_\pi^2}} dq
\end{equation}
has the VPI value
$
{\cal J} = -1.041 \mbox{ mb},
$
we find
\begin{equation}
{\displaystyle f_{\pi NN}^2 \over \displaystyle 4 \pi} = 0.0766,
\qquad \mbox{or} \qquad
{\displaystyle g_{\pi NN}^2 \over \displaystyle 4 \pi} = 13.84,
\end{equation}
which is fully consistent with the value of $13.75 \pm 0.15$ given in 
\cite{arnpin}.

\subsection{$\pi^-p$ Atoms}

The Coulomb attraction between $\pi^-$ and $p$ causes a
bound system called pionic hydrogen \PH. Additionally, the hadronic 
interaction between
the two constituents distorts the short range interaction and 
changes the pure Coulomb spectrum.
Decay channels
$ \pi^- p \to \pi^0 n$ and $ \pi^- p \to \gamma n$
allow a rapid decay.

The hadronic shift of the $3p \to 1s$ transition and the total $1s$ width has 
recently been measured at PSI \cite{sigg95}. To analyze this experiment we
use quantum inversion for the determination of the hadronic potentials.
Inelasticities, Coulomb and other isospin breaking effects are 
supposed to be not included in the SM95 phase shift analysis \cite{arnpin}.
Using the  $S_{11}$ and $S_{31}$ phase shifts shown in
Fig. \ref{figure1}, we obtain the hadronic potentials $V_{11} (r)$ and
$V_{31} (r)$, Fig. \ref{figure2} (left). The inversion algorithm 
uses only the nonrelativistic reduced mass $\mu = 121.50$ MeV 
based upon the 
$\pi^\pm$ and $p$ masses, consistently with the phase shift analysis.  

In the next step  we apply a rotation of these potentials in isospin space
from good isospin states into particle eigenstates. This yields the
potential matrix $V^{ij} (r)$
of the coupled radial Schr\"odinger equation 
\begin{equation} \label{pinbeqn1}
f_i^{\prime\prime} + k_i^2 f_i = \sum_{j=1}^2 2 \mu_i V^{ij} f_j,
\quad i=1(2)\mbox{\ \ for\ \ }
\pi^-p \ (\pi^0n)
\end{equation}
for the $\pi^- p \leftrightarrow \pi^0 n$ system.
The potential matrix contains the isospin rotation coefficients and
the Coulomb attraction
\begin{eqnarray}
 V^{11}  & = &  V^{\pi^-p} =
\frac23 V_{11} + \frac13 V_{31} - \frac{e^2}{r}, \\
V^{12}  = V^{21}  & = & -{\sqrt{2}\over 3}(V_{11} - V_{31}), \\
V^{22}  & = & V^{\pi^0n} = \frac23 V_{31} + \frac13 V_{11}.
\end{eqnarray}
The hadronic part of this matrix is shown on the right side of 
Fig. \ref{figure2}. The reduced masses we used are $\mu_1= 121.50$ MeV and 
$\mu_2= 118.02$ MeV. 
In Table \ref{tab_pinb1s} we signal this choice with the parameter $x=-1$.
This mass difference introduces an additional isospin breaking effect.
Since this issue is of central importance for the pion--pion system in the 
next section we investigate this point in further details also here
and distinguish the alternative calculation with $\mu_1= \mu_2= 121.50$ MeV
and $x=+1$.
Generally it is assumed that charge and mass isospin--breaking effects should 
have a small impact on the 
lifetime and width of \PH\, or the eigenchannel phase shifts.
  
The bound states of the  $\pi^-p$ system can be found as resonances in the 
energetically open $\pi^0n$ channel, showing a partial width 
$\Gamma_{ns}^{\pi^-p\to\pi^0n}$ deduced from the elastic cross section  
\begin{equation}
\sigma(\pi^0n \to \pi^0n)={\pi\over k_2^2}|1-S_{22}|^2.
\end{equation}
Results for the ground state are given in Table 
\ref{tab_pinb1s}. 
To account for the full experimental width
one has to correct the partial width with the 
Panofsky ratio $P =1.546 \pm 0.009$ \cite{sigg95}
\begin{equation}
\Gamma_{ns} = \left(
1 + \frac1P \right) \Gamma_{ns}^{\pi^-p\to\pi^0n}.
\end{equation}
To show the effect of isospin breaking caused by different reduced masses in
channel 1 and 2, we repeated this calculation with $\mu_2 (\pi^0n)$ set equal to
$ \mu_1 (\pi^-p)$. These results are also included in Table \ref{tab_pinb1s}
with $x=+1$.
Furthermore we have accounted for the $\gamma n$ decay by the 
introduction of a phenomenologic imaginary Gaussian potential $W(r) = W_{11} =
-9 \exp(-4r^2)$ in eqn. (\ref{pinbeqn1}), where we replace $V_{11} \to
V_{11} +i W_{11}$. The point Coulomb potential is replaced by 
double folded Gaussian charge distributions for the pion and nucleon
\[ 
{e^2 \over r} \to {e^2 \Phi(1.13 \; r) \over r} .
\] 
All results are shown in Table \ref{tab_pinb1s}. They agree well with the
experimental values, and the small isospin breaking effects 
from the mass difference 
confirm our assumption and establish an excellent support of the 
inversion approach.

\section{$\pi\pi$ Scattering}
$\pi\pi$ phase shifts come from the analysis of final state interactions in
$
\pi N \rightarrow \pi \pi N
$
systems or the $K_{e4}$--decay   
$
K^- \rightarrow \pi^+ \pi^- e \bar{\nu}.
$
Here we use results of the CERN--Munich experiment \cite{fro77}
$
\pi^- p \to \pi^+ \pi^- n 
$
for all channels with $\ell \leq 3$. 
The scattering is purely elastic up to $M_{\pi\pi} = 987.3$ MeV where
a coupling $ \pi^+ \pi^- \leftrightarrow K \bar{K} $
becomes dominant.
In addition to the experimental data of Froggatt et al. \cite{fro77}
we use data from meson exchange models \cite{loh90} and 
chiral perturbation theory $\chi$PT \cite{gas83}.
Notation for the isospin and angular momentum channels uses 
$\delta_\ell^T \mbox{or } V_\ell^T$.
The inversion results are shown in Fig.\ 4.  

Similar to the $\pi^- p$ system there exists pionium \PP\
which is formed by $\pi^- \pi^+$ Coulomb attraction and decays 
by charge exchange into the open $\pi^0 \pi^0$ channel.  
The coupled channel system is equivalent to eqn. (12) replacing $p \to \pi^+$
and $n \to \pi^0$.
We assume the same approach and rotate the good isospin potentials into
particle states \cite{pionium}
which reduces to the same form as given by eqn. (13) -- (15), simply
replace $V_{11} \to V_0^0$ and $V_{31} \to V_0^2$. 
Phase shift analyses and inversion use a single mass $\mu = m_{\pi^+}/2$
without Coulomb effects. This assumtion guarantees good 
isospin $T=0$ and 2. With the purpose to display uncertainties in the 
phase shifts we used three sources and distinguish a set of three potentials
for $V_0^0$ and $V_0^2$ respectively. They are shown in Fig.\ 4 (right), of
which the $V_0^0$ potential is of particular interest. Similar to the
$\pi N$ $P_{33}$ channel potential we find here a potential barrier and a
very strong short range attraction. Such potential may be able to
support a potential resonance similar to the $\Delta$--resonance in the
$\pi N$ $P_{33}$ channel, but here in the $\ell=0, T=0$ channel the resonance
width is expected larger than for $\ell=1$ since the resonance conditions
are more delicate due to the centrifugal barrier. The radial dimensions and
potential strenghts are quite comparable. It is obvious to ask if the implied 
great width resonance supported by this potential can be identified with the
isoscalar $\sigma$ meson which OBE potential require
and which is of general interest 
[23,24,25,26]. It is intended to
study this resonance in more details by our OSBEP approach and use also other 
experimental information.
A first glance on this investigation is given by simply using different 
reduced masses in eqn. (12). This changes effectively the strength of $V_0^0$
within a few \% but causes a dramatic change of the eigenchannel 
phase shifts which can be identified with the $T=0$ isospin state. In 
Fig. 5 we show three cases of different choices of $\mu_i$. Fig. 5
(top left) $\mu_1 = \mu_2 = m_{\pi^+}/2$ without Coulomb potential. This
confirmes the original input phase shifts used for the inversion. 
(top right) $\mu_1 = m_{\pi^+}/2$, $\mu_2 = m_{\pi^0}/2$ which are the 
correct reduced masses in the nonrelativistic limit. As expected, the 
eigenchannel phase $\delta_0^2$ remains unchanged, whereas
$\delta_0^0$ is now far off the data. (bottom) 
$\mu_1 = \mu_2 = m_{\pi^0}/2$ is a further step, and we observe 
little change for $\delta_0^2$ but a dramatic effect for $\delta_0^0$.
This investigation is supposed to show that the strength of the potential
enters very sensitively and changes the $\delta_0^0$ phase in a wide range 
quite untypical for a resonance. The effects upon the lifetime of
pionium has already been studied and we observe a dramatic change of 
lifetime compared between what is given in Table IV and \cite{pionium}. 
This change requires a deeper 
understanding since the hypothetical $\sigma$--meson is producing the 
medium--range $NN$  attraction. Medium effects of this resonance
should be of particular importance and we are presently studying this aspect
in microscopic optical potentials for nucleon--nucleus scattering.

\section{$K^+N$, $K\pi$ and $K\bar{K}$ Scattering}
\subsection{$K^+N$ Scattering}
Kaons introduce strangeness into the inversion algorithm where the
configurations $K^+N, K^0N$ with $S=+1$ and $K^-N, \bar{K}^0N$
with $S=-1$ are allowed in $KN$ scattering.
The $S=-1$ channels show strong resonance effects whereas the the $S=+1$
channels are smooth. 
Phase shift analyses are restricted by Arndt in SP92 \cite{arnkn}
to the $K^+N$ system.
In Fig. \ref{abb_kn1} we show the result for $\ell=3$, the nomenclature used is
$\ell_{T,2J}$, $J=\ell \pm \frac12$.
The results show little structure in the radial dependencies.
Undoubtedly the range of the potential is longer than for the $\pi N$ system. 
The long range part of the interaction is discussed in section V.
We calculated $K^+N$ s--wave scattering lengths from  inversion potentials
and compare them with predictions of the phase shift analysis and models.
They are summarized in Table \ref{tab_kns}. Predictions
from meson exchange models \cite{bue90} agree either with the $T=0$ or $1$
scattering lengths.

\subsection{$K\pi$ scattering}
Suitable informations for $K\pi$ scattering come from final state interaction
analyses of the reactions
$K^{\pm} p  \to  K^{\pm} \pi^+ n, $ and $ K^{\pm} p \to K^{\pm} \pi^- 
\Delta^{++}$.
The notation for channel identification is 
$ \delta_\ell^{2T} $,
where $T=\frac12$ or $\frac32$.
Well known resonances in this system  are $K^\ast (892)$ and $K^\ast_0 (1430)$.
The phase shift analysis starts at $M_{K\pi} = 0.73$ GeV 
and remains elastic up to 1.3 GeV \cite{est78}. 
There is a gap between threshold, $M_{K\pi} = 0.63$ GeV, 
and the first data points.
We bridged this gap with 
a smooth extrapolation with 
$\lim_{k\to 0} \delta_\ell^{2T} (k) = {\cal O} (k^{2\ell+1})$
\cite{san96}. 

Our scattering lengths, see Table \ref{tab_piks}, 
agree well with estimates from dispersion relations
($0.22 \geq a_0^1 \geq 0.045$, $-0.10 \geq a_0^3 \geq -0.165$) \cite{kar80}
and the experimental values \cite{est78}, but they are 1.4 to 3 times 
larger than predictions from various models. To solve this
puzzle, more data between the $K\pi$ threshold and $M_{K\pi} = 730$ MeV
are needed.

\subsection{$K\bar{K}$ Scattering}
There exist no phase shift analyses for this system, and we have to rely upon 
an effective range expansion by
Kaminski and Lesniak \cite{kam94}. Their expansion may be
disputed since they neglect inelasticities. This describes 
briefly the experimental situation with the implication that our
analysis represents only some qualitative features.

We are using two sets of parameters from \cite{kam94} in
the effective range expansion
\begin{equation} \label{kkbar_er}
k \cot \delta_K (k) = \frac{1}{\mbox{Re } a_K} +
\frac12 R_K k^2 + V_K k^4 
\end{equation}
whose values are given in Table \ref{tab_kkbars}.
The phase shifts $\delta_0 (M_{KK})$ from this parameterization 
are shown in Fig. \ref{abb_kkbar1}, together with 
the inversion potentials which reproduce the 
effective range expansion with high precision.
There exists a claim from lattice QCD that the short range attractive 
interaction has an explanation in a non--vanishing propagator structure 
\cite{fie95}. 

\section{Long Range Behaviors}

In the display of inversion potentials we restricted ourselves to the big 
effects at distances between 0--2 fm and emphasized the short range domain.
Actually, the most reliable information is the long range part of the
potentials which can be parameterized in terms of a 
Yukawa potential with a range parameter given by the Compton wave length of
an exchanged particle. For the pion--nucleon system we have also dwelled upon 
the coupling constant. In Table \ref{tab_yukvgl}
 are summarized the long range
Yukawa parameters 
$
V(r) = Y e^{-\mu r} / r
$
which we extracted from our inversion potentials in the $\ell = 0$ channels. 
We find that the exchanged masses in $pp$, $K\bar{K}$ and $K^+N$
may be interpreted as 
 one--pion, two--pion and one--$\sigma_1$ exchange respectively. From the
microscopic point of view, the propagators of a dominant
pseudoscalar or scalar s--channel exchange transform into a Yukawa--like
potential tail. In $\pi N$, $\pi K$ and $\pi\pi$ scattering this
interpretation is not valid, since here s-- and t--channel graphs
contribute and thus a transformation into coordinate space does not lead
to a Yukawa with a physical mass of an exchange particle.
A more detailed discussion of this table and its implications can be 
found in \cite{san96}.

\section{Summary and Conclusions}
With this contribution we show that the rich sources of phase shift analyses 
for general hadronic systems can successfully be used to obtain a
qualitative and sometimes quantitative understanding of the interaction
in terms of a simple local potential. 
The ranges and strengths of these potentials are often determined
by the masses of the scattered particles and the spin--isospin dependence
of the partial waves. This dependence can often be understood in terms 
of a boson exchange picture and ultimately may be related to the underlying
QCD dynamics. The latter aspect is most obvious in the $p$--wave resonances
of the $\pi N$ system. With this first attempt of using quantum inversion 
to study the realm of hadronic interactions we establish encouragement
to look for alternative equations of motion which should account better
for the relativistic kinematics which is definitely important for the
lighter hadronic systems or any extension towards higher energy.
Various relativistic wave equations have been studied and
applied in recent years when treating hadrons with quarks as their 
constituents. 
\begin{acknowledgments}
Supported in part by Forschungszentrum J\"ulich, COSY Collaboration
41126865.
\end{acknowledgments}

%
%

\begin{table}\centering
\caption[$\pi N$ Resonanzparameter aus der Inversion]%
{$\pi N$ resonance parameters used in the inversion scheme.}
\label{tab_pin2}
\begin{tabular}[thb]{cccccc}
\str $\pi N$ channel & Re($k_r$) [fm$^{-1}$] & Im($k_r$) [fm$^{-1}$] &
Mass [MeV] & Width [MeV] & Name \\[0.1cm]
\hline
\str  $P_{11}$ & 1.8200 & $-0.6200$ & 1381 & 159 & N(1440) \\
 $P_{33}$ & 1.0665 & $-0.2440$ & 1212 & 102 & $\Delta$(1232) \\
 $D_{13}$ & 2.2900 & $-0.1050$ & 1514 &  93 & N(1520) \\
 $F_{15}$ & 2.8520 & $-0.1230$ & 1674 &  72 & N(1680) \\
\end{tabular}
\end{table}

\begin{table}\centering
\caption[$\pi N$ s--Wellen Streul\"angen]%
{$\pi N$ s--wave scattering lengths from several models and
experimental analyses
together with predictions for the $\pi NN$ coupling constant
obtained from the GMO sum rule.}
\label{tab_pins}
\begin{tabular}[htb]{cccccccccc}
\str & Model & \hspace{0.4cm} & $a_1$ [$m_\pi^{-1}$]  & \hspace{0.4cm} & $a_3$
[$m_\pi^{-1}$]  & \hspace{0.4cm} & $f_{\pi NN}^2/4\pi$ & \hspace{0.4cm} &
  Ref. \\[0.1cm]
\hline
\str &  SM95 Inversion & & 0.178  & & $-0.088$ & & 0.0766& &  \\
\str & KH80 & & 0.173 & & $-0.101$ & & 0.079 & & \protect\cite{koch80} \\
& $\pi^-p$ 1s state & &  0.185 & &  $-0.104 $ & & 0.081 & &
\protect\cite{sigg95} \\
\str & Pearce et al.& & 0.151 & & $-0.092$ & & 0.072 & &
\protect\cite{pea91} \\
 & Sch\"utz et al. & & 0.169 & & $-0.085$  & &  0.074 & &
\protect\cite{sch94a} \\
\end{tabular}
\end{table}

\begin{table}\centering
\caption{Hadronic shift of the $1s$ level, with respect to the pure Coulombic
reference energy $E_{1s}^C= 3234.9408$ [eV], and partial widths $\Gamma^{\pi^-p
\to \pi^0n}_{1s}$.
These partial widths have to be multiplied with 1.647 to account
for the Panofsky ratio to obtain the total widths.
This yields a typical value $\sim 0.862$ [eV] to be compared
with the experimental value
$\Gamma_{1s}$ = 0.97$\pm$0.10$\pm$0.05 [eV].
The strength of the imaginary part in the last entry $W_{11}$ is adjusted to
reproduce the experimental value.}
\label{tab_pinb1s}
\begin{tabular}[b]{cccccccc}
 \str & \hspace{0.6cm} &\multicolumn{2}{c}{Point Charge Coulomb}&
\hspace{0.6cm} & \multicolumn{2}{c}{Gaussian Charge Coulomb}&
\\   
\str x & & Shift [eV] & FWHM [eV] & & Shift [eV] & FWHM [eV] &  \\
\hline
\str --1 & &  --7.13259 & 0.5187 & &   --7.01055 & 0.5144  &   \\
+1 & & --7.29821 & 0.5250 & &  --7.16746 & 0.5230  &    \\
\strr $\mu(s)$ & & --7.26334 & 0.5317 & & --7.13259 & 0.5266  &     \\
\multicolumn{7}{c}{ Shift =
--7.127$\pm$0.046  [eV],
$\Gamma_{1s}^{\pi^0n}$ = 0.590 [eV] }
& Sigg \cite{sigg95}
\\
\strr --1 & & --7.23259 & 0.9763  & & --7.12387 & 0.9763 & $V_{11}+iW_{11}$
\\
\multicolumn{7}{c}{ Shift =
--7.127$\pm$0.028$\pm$0.036  [eV],
$\Gamma_{1s}$ = 0.97$\pm$0.10$\pm$0.05 [eV]}
& Sigg \cite{sigg95} \\
\end{tabular}
\end{table}

\begin{table}\centering
\caption{\PP properties from inversion potentials.
The Point Coulomb ground state reference energy $E_{1s}^C= 1.85807248$ [keV].
(x=+1) $\equiv \mu_1 = \mu_2 = m_{\pi^+}/2.$ }
\label{table3}
\begin{tabular}[b]{ccccccccccc}
\str   x & \hspace{0.3cm} & $E_{1s}$ [keV] &\hspace{0.2cm} &
\multicolumn{1}{c}{Shift  [eV]}  &\hspace{0.2cm} &
\multicolumn{1}{c}{$\tau$ [$ 10^{-15}$ sec]} &\hspace{0.2cm} & FWHM  [eV]
&  & Reference \\
\hline
\str +1& &  1.8638814  & &--5.809  & & 1.97   & & 0.3481 & &  Froggatt\\
 $\mu(s)$ & & 1.8636070  & & --5.538 & & 2.05& &0.3337 &  & Froggatt \\
\str +1& &  1.8635114 & &--5.439 & &  1.89  & & 0.3627 & &  Lohse\\
 $\mu(s)$ & & 1.8632880  & & --5.216 & & 2.03 & & 0.3385 & & Lohse\\
\str +1& &  1.8616174  & &--3.545 & & 3.22  & &0.2128 & &  $\chi_{PT}$\\
 $\mu(s)$ & & 1.8614390  & & --3.367 & & 3.37 &  &0.2031& & $\chi_{PT}$\\
\multicolumn{11}{c}{\em  \str Predictions from experimental analysis
and other models} \\
\str  & &  1.858 & &   & &  2.90$^{+\infty}_{-2.10}$ & &  & & Afanasyev
\protect\cite{afa93} \\
   & &1.865 & & --7.0    & &  3.20 & & & & Efimov  \protect\cite{efimov}
\\
\end{tabular}
\end{table}

\begin{table}\centering
\caption[$K^+N$ s--Wellen Streul\"angen]%
{$K^+N$ s--wave scattering lengths from several models and
experiment.}
\label{tab_kns}
\vspace*{0.2cm}
\begin{tabular}[tbh]{ccccccc}
\str Modell & \hspace{0.4cm} & $a_0$ $(T=0)$ [fm]  & \hspace{0.4cm} &
$a_0$ $(T=1)$ [fm]  & \hspace{0.4cm} &   Ref. \\[0.1cm]
\hline
\str SP92 & & 0.00 & & $-0.33$ & &  \protect\cite{arnkn} \\
Inversion & & 0.00 & & $-0.33$ & &  \\
\str Meson Ex. (A)&  & 0.03 & & $-0.26$ & &  \protect\cite{bue90} \\
Meson Ex. (B1)& & $-0.15$ & & $-0.32$ & &  \protect\cite{bue90} \\
\end{tabular}
\end{table}

\begin{table}[thb]\centering
\caption[K$\pi$ s--Wellen Streul\"angen]%
{K$\pi$ s--wave scattering lengths from different theoretical models and
experimental analyses.}
\label{tab_piks}
\begin{tabular}[h]{ccccccc}
\str Model & \hspace{0.6cm} & $a_0^1$ [$m_\pi^{-1}$]  & \hspace{0.6cm} &
$a_0^3$ [$m_\pi^{-1}$]  & \hspace{0.6cm} &   Ref. \\[0.1cm]
\hline
\str Estabrooks & & 0.331 &  & $-0.138$ & &  \protect\cite{est78} \\
Inversion & & 0.340 & & $-0.147$ & &  \\
\strr Meson Ex.& & 0.23 & & $-0.064$ & &  \protect\cite{loh90} \\
 $\chi$PT & & 0.17 & & $-0.05$  & &  \protect\cite{ber91a} \\
 Quark model & & 0.23 & & $-0.077$  & &  \protect\cite{bar92,li94} \\
\end{tabular}
\end{table}

\begin{table}[th]\centering
\caption%
{$K\bar{K}$ isoscalar scattering lengths from \protect\cite{kam94}
and their reproduction by inversion potentials.}
\label{tab_kkbars}
\begin{tabular}[b]{ccccccc}
\str  Modell & \hspace{0.6cm} & Re $a_K$ [fm] & \hspace{0.6cm} & $R_K$ [fm]
& \hspace{0.6cm} & $V_K$ [fm$^3$] \\[0.1cm]
\hline
\str Set 1 & & $-1.73$ & & $0.38$ & & $-0.66$  \\
Inversion & & $-1.73$ & & $0.38$ & &   \\
\str Set 2 & & $-1.58$ & & $0.20$ & & $-0.83$  \\
Inversion & & $-1.58$ & & $0.20$ & &   \\
\end{tabular}
\end{table}

\begin{table}\centering
\caption[Yukawa--Verhalten verschiedener Streuungen]%
{Yukawa parameters. The mass of the exchanged particle is given by
$\mu = \frac{mc}{\hbar}$.}
\label{tab_yukvgl}
\begin{tabular}[b]{ccccccc}
\str   & \hspace{0.6cm} & $Y$ [MeVfm] & \hspace{0.6cm} & $\mu$ [fm$^{-1}$]
& \hspace{0.6cm} & $m$ [MeV]
\\[0.1cm]
\hline
 $pp$       & & 14.4    & & 0.684  & &  134.97 \\
 $K^+N$     & & 1325.85 & & 2.9436 & & 580.85 \\
 $K\bar{K}$ & & 1923.59 & & 1.316  & & 259.68 \\
 $\pi N$    & & 111.84  & & 1.61   & & 317.70 \\
 $\pi K$    & & 638.1   & & 1.87   & & 368.4 \\
 $\pi\pi$   & & 1081.28 & & 2.219  & & 437.87 \\
\end{tabular}
\end{table}

\unitlength 1cm
\begin{figure}[hbt]\centering
\begin{picture}(8,8)(0.0,0.0)\centering
\epsfig{figure=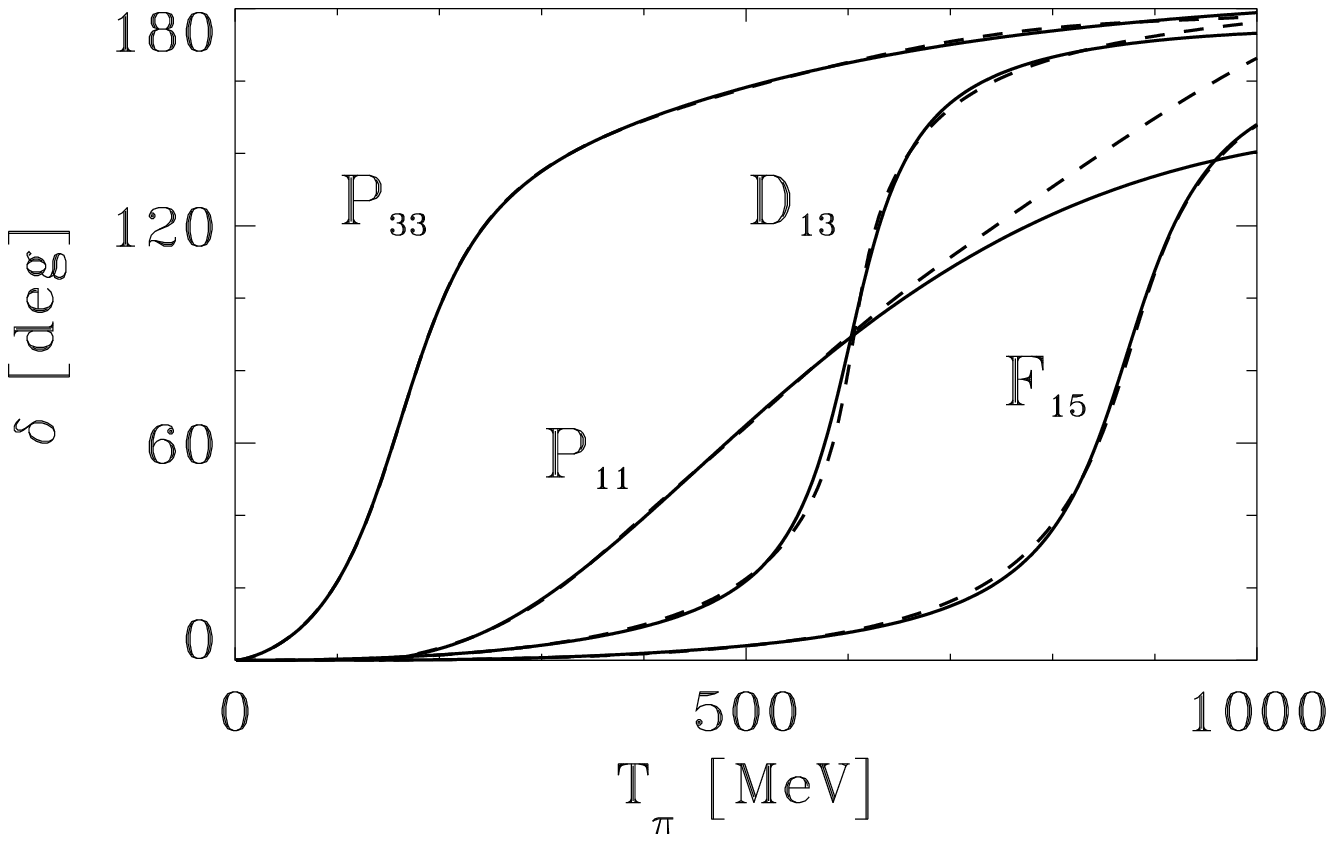,width=8cm}
\end{picture}
\begin{picture}(8,8)(0.0,0.0)\centering
\epsfig{figure=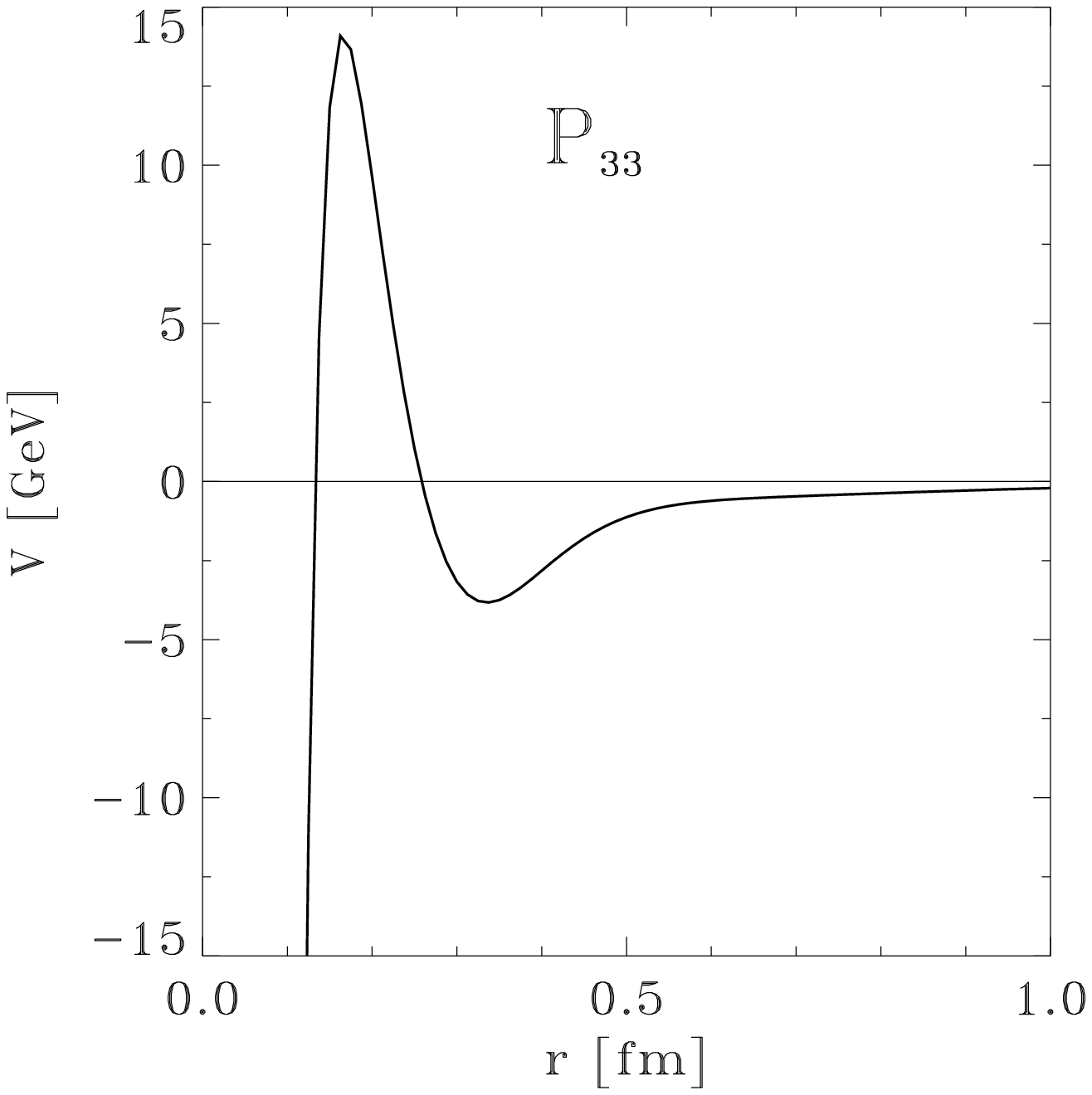,width=8cm}
\end{picture}
\caption%
{Left: $\pi N$ real phase shifts in the resonant channels
(dashed)
and their
reproduction by the inversion potentials (full line).
Right: $\pi N$ $P_{33}$ potential from quantum inversion.}
\label{abb_pin1}
\end{figure}

\begin{figure}\centering
\begin{picture}(8.0,8.0)(0.0,0.0)
\epsfig{figure=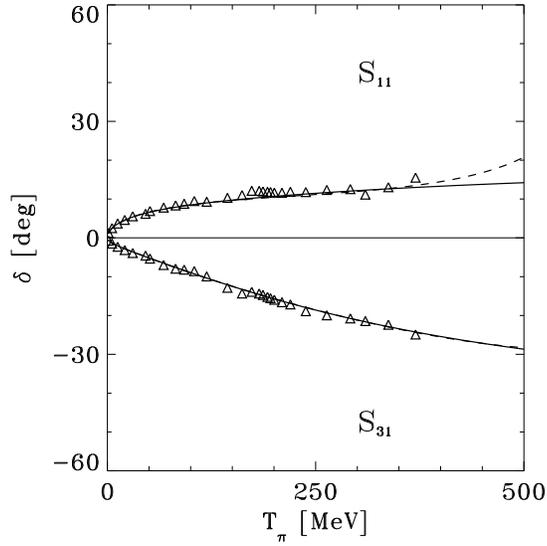,width=8.0cm}
\end{picture}
\caption{$\pi N$ SM95 \protect\cite{arnpin} (dashed line) and
KH80  \protect\cite{koch80} (triangles) data with
reproduction  by inversion potentials  (solid line) for the
$\pi N$ $S_{11}$ and $S_{31}$ channels.}
\label{figure1}
\end{figure}

\begin{figure}\centering
\begin{picture}(8.0,8.0)(0.0,0.0)
\epsfig{figure=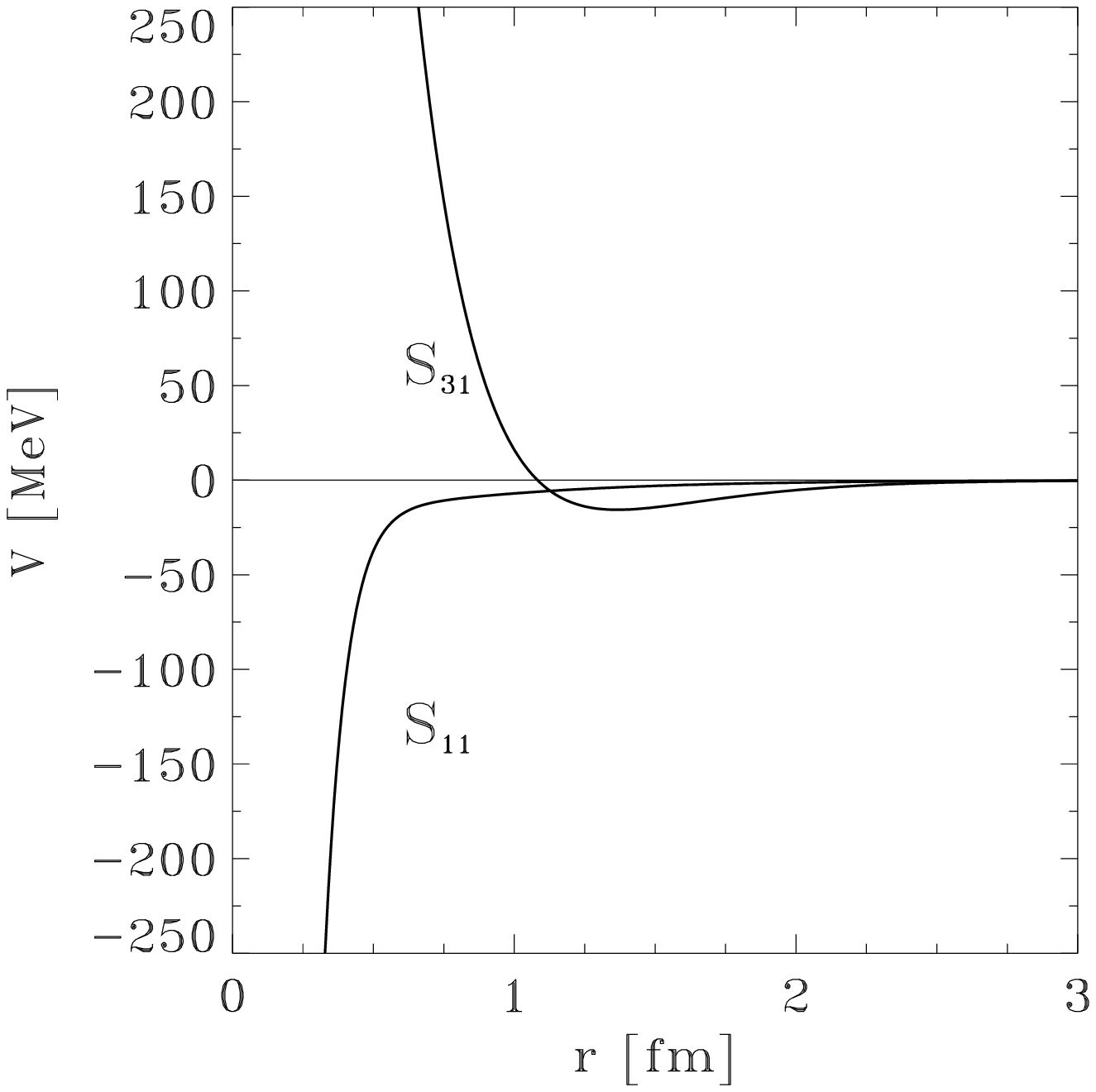,width=8.0cm}
\end{picture}
\begin{picture}(8.0,8.0)(0.0,0.0)
\epsfig{figure=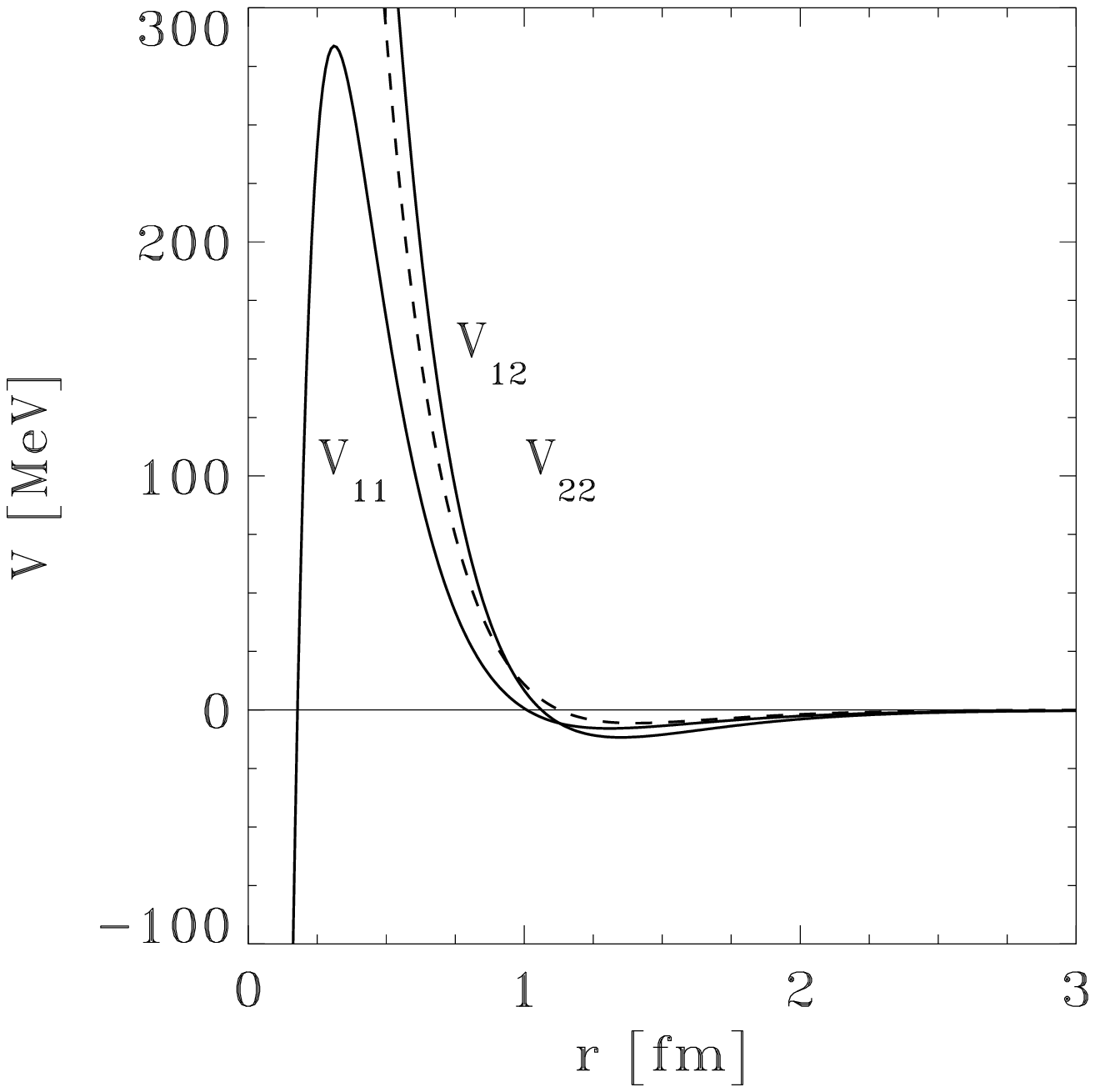,width=8.0cm}
\end{picture}
\caption%
{$\pi N$ inversion potentials using SM95 phase shifts
(left), and the potential matrix (right).}
\label{figure2}
\end{figure}

\unitlength 1cm
\begin{figure}[ht]\centering
\begin{picture}(8,8)(0.0,0.0)\centering
\epsfig{figure=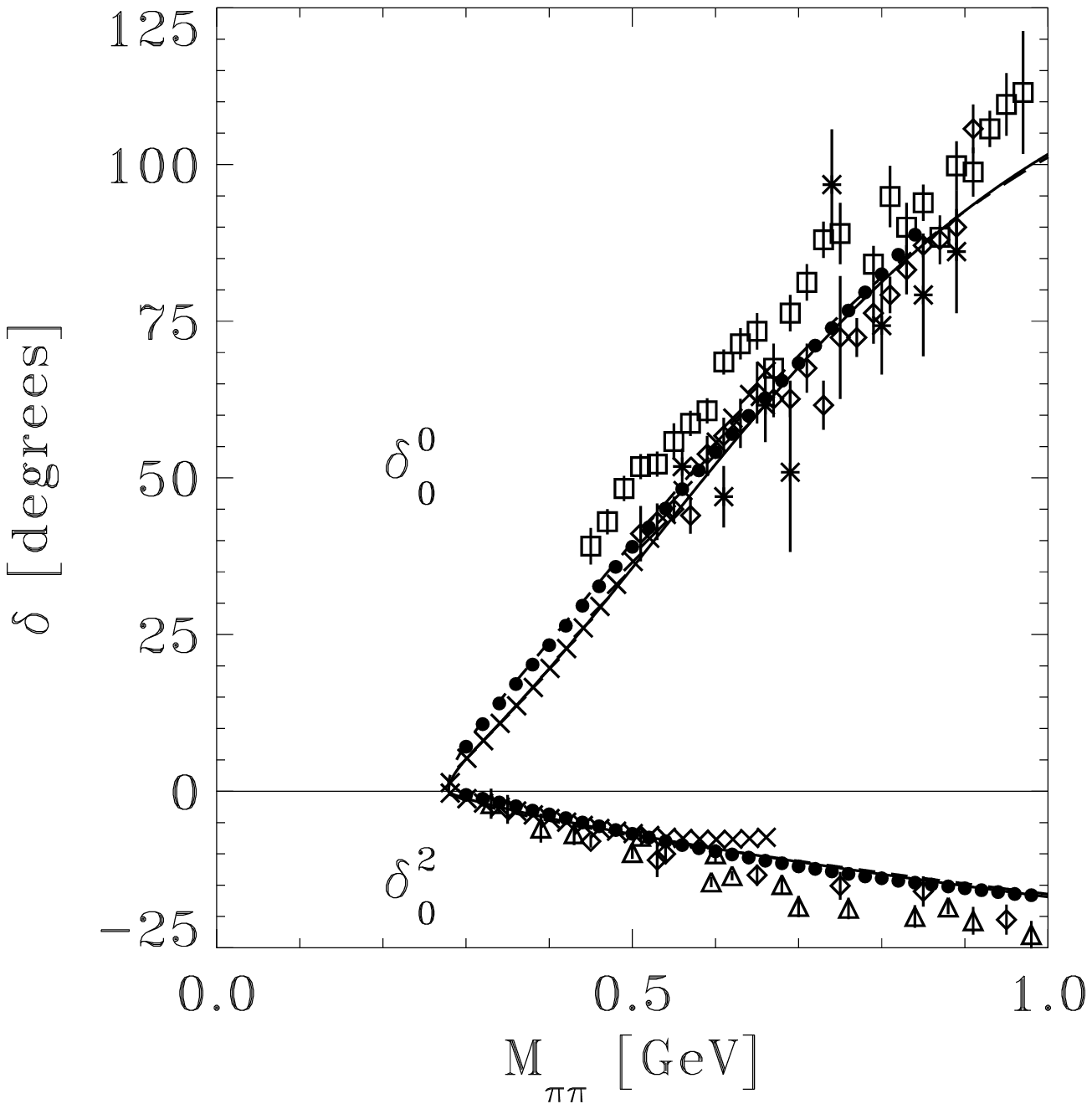,width=8cm}
\end{picture}
\begin{picture}(8,8)(0.0,0.0)\centering
\epsfig{figure=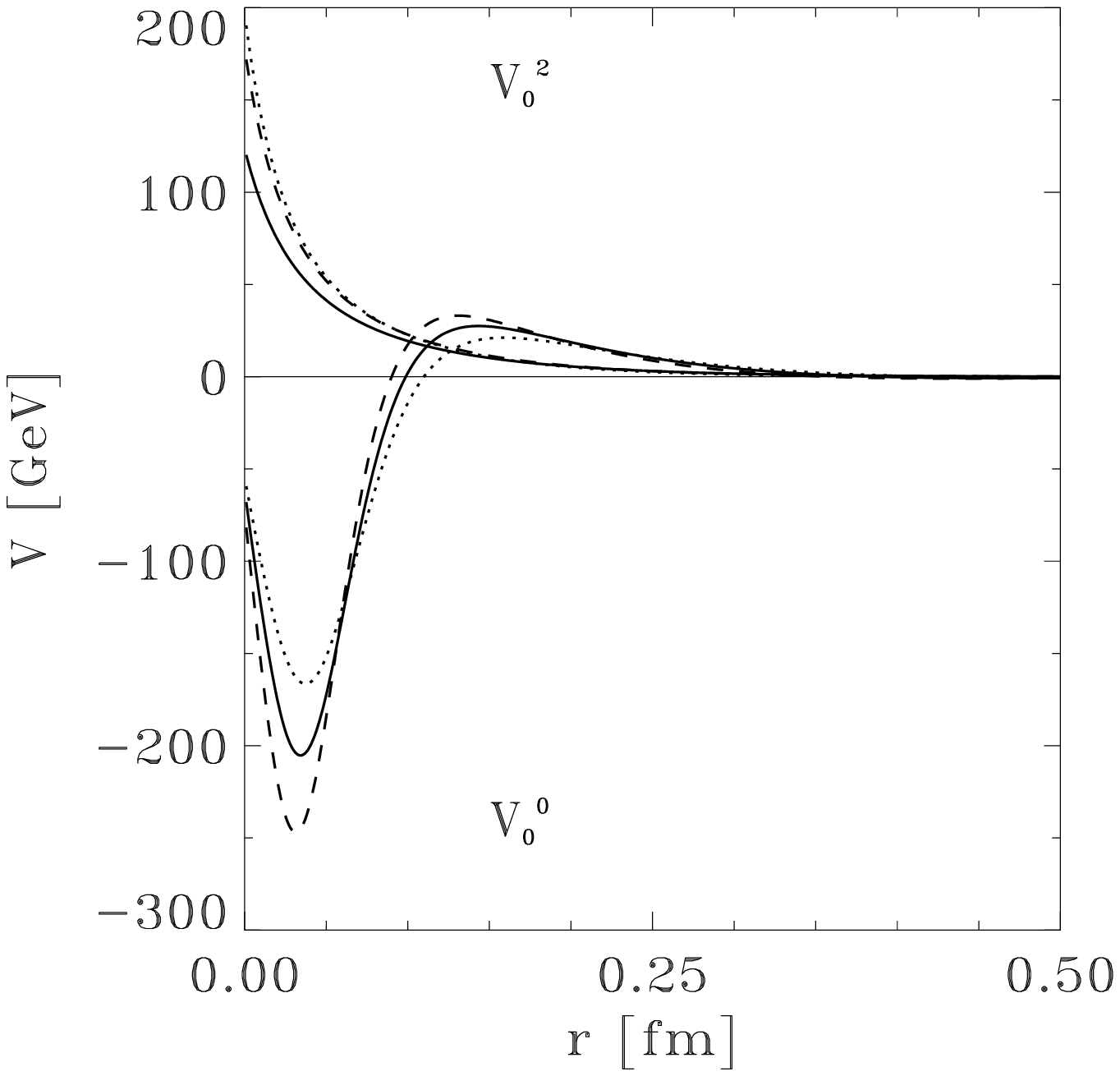,width=8cm}
\end{picture}
\caption%
{Left: $\pi\pi$ $\ell=0$ phase shifts from
$\chi$PT (crosses) and the reproduction by the inversion potentials (full line),
from the analysis by Froggatt (dots) and
the reproduction by the inversion potentials (dashed).
Right:
$\pi\pi$ $\ell=0$ inversion potentials based on phase shifts from $\chi$PT
(full line), the analysis by Froggatt (dashed) and meson exchange (dotted).}
\label{abb_pipi1}
\end{figure}

\unitlength 1cm
\begin{figure}[ht]\centering
\begin{picture}(8.0,8)(0.0,0.0)\centering
\epsfig{figure=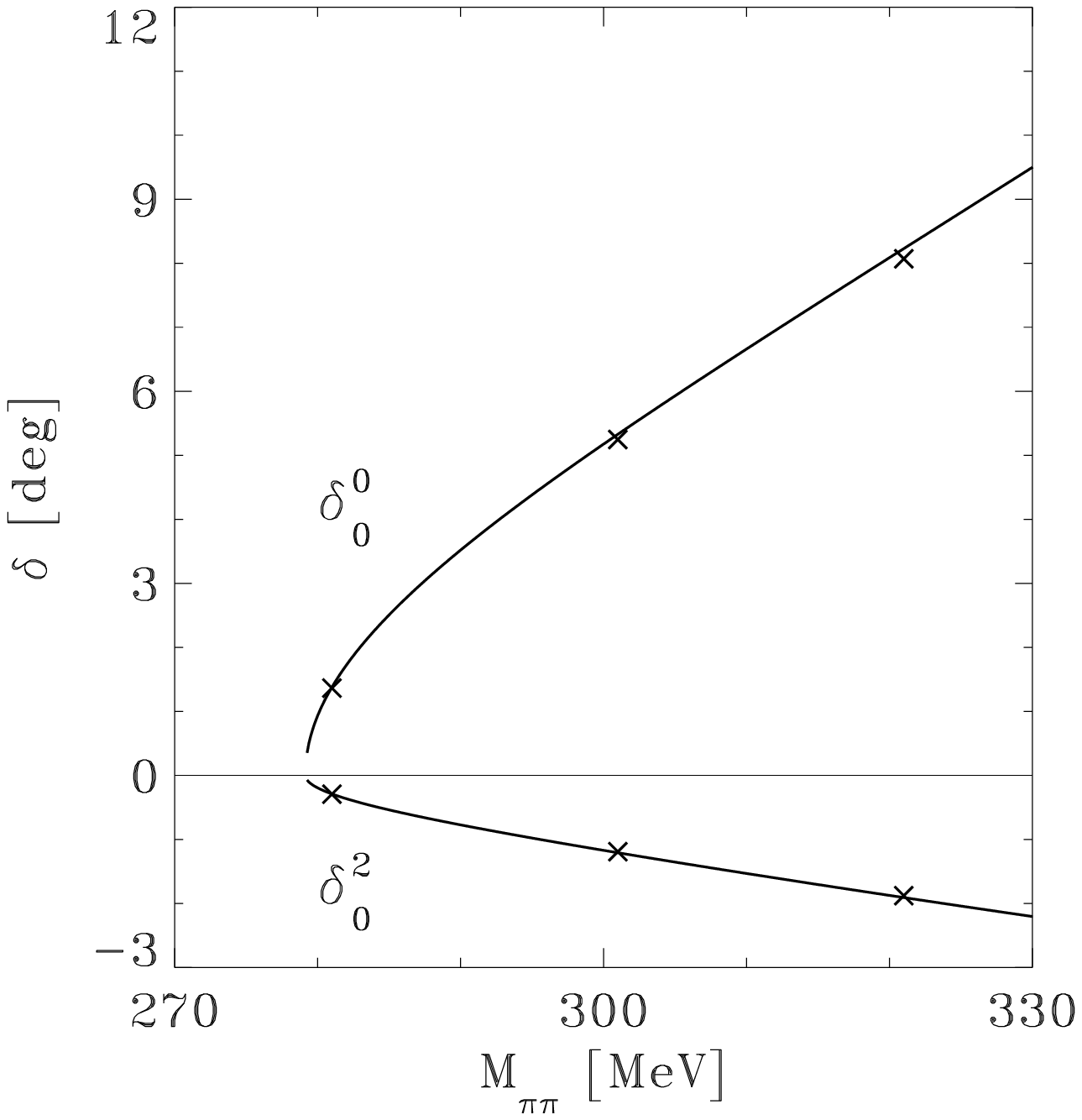,width=8cm}
\end{picture}
\begin{picture}(8,8)(0.0,0.0)\centering
\epsfig{figure=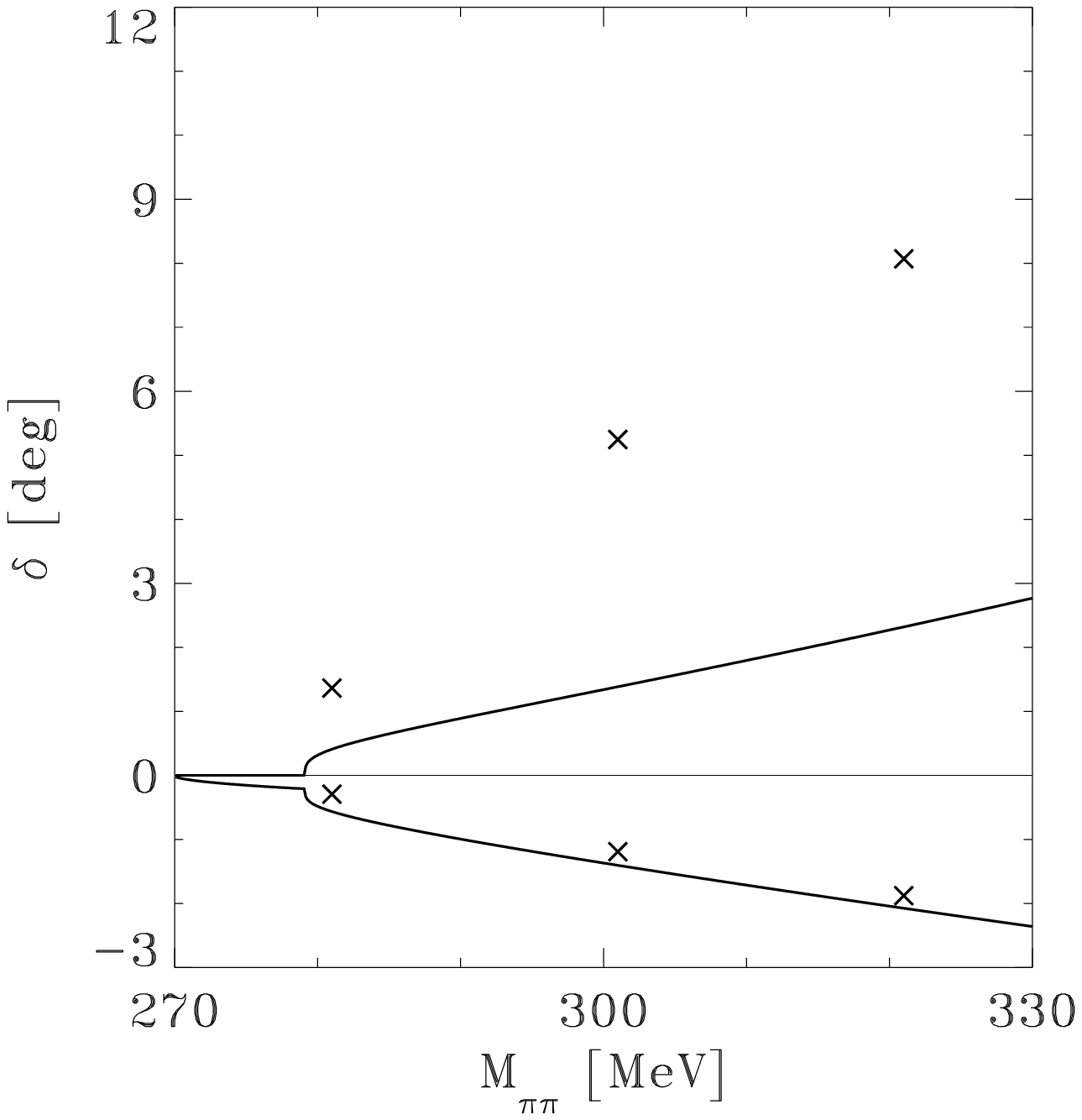,width=8cm}
\end{picture}
\begin{picture}(8,8)(0.0,0.0)\centering
\epsfig{figure=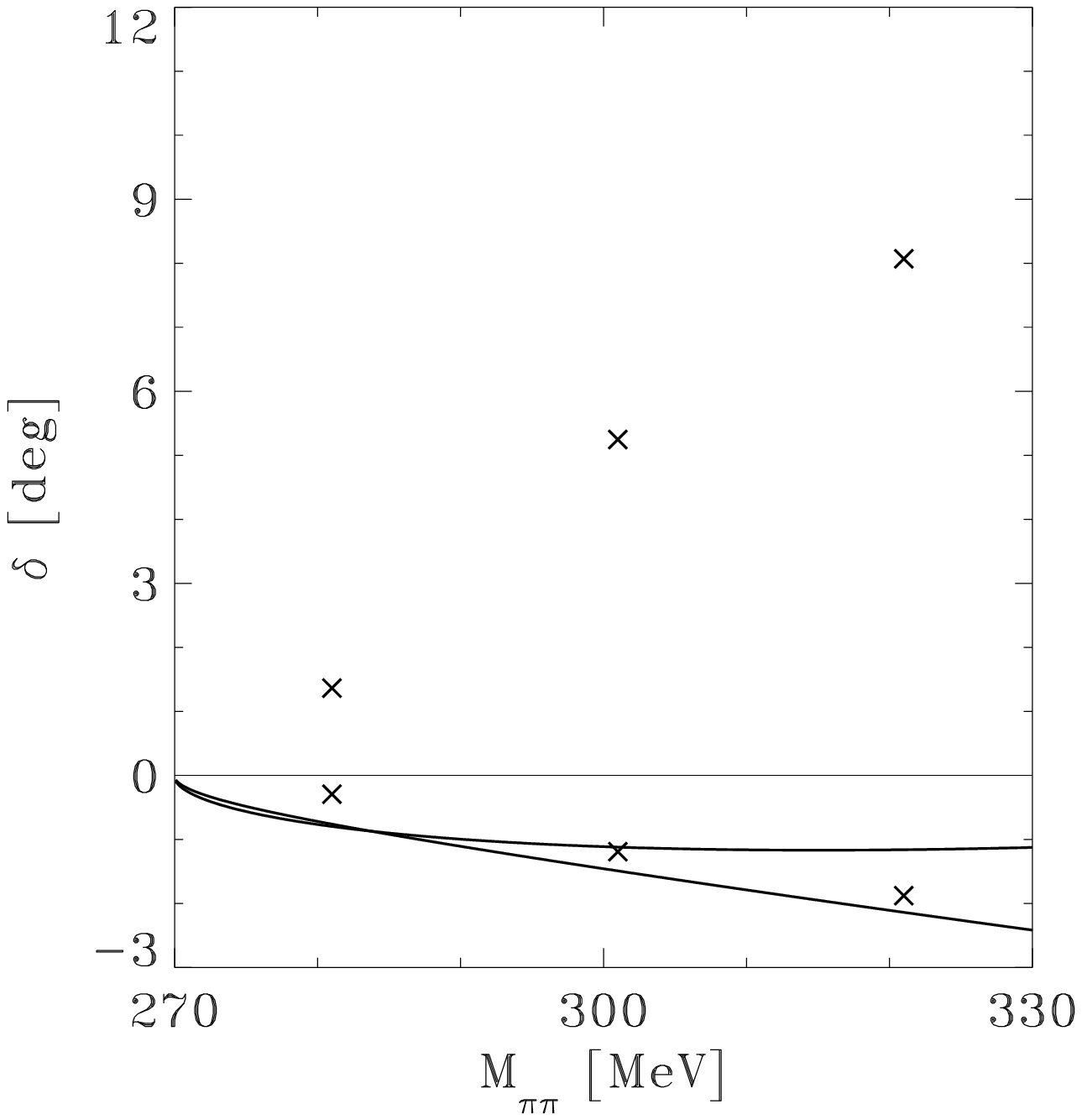,width=8cm}
\end{picture}
\caption%
{Eigenchannel phase shifts. More details given in the text.}
\label{abb_pipipha}
\end{figure}

\unitlength 1cm
\begin{figure}[bht]\centering
\begin{picture}(8.0,8)(0.0,0.0)\centering
\epsfig{figure=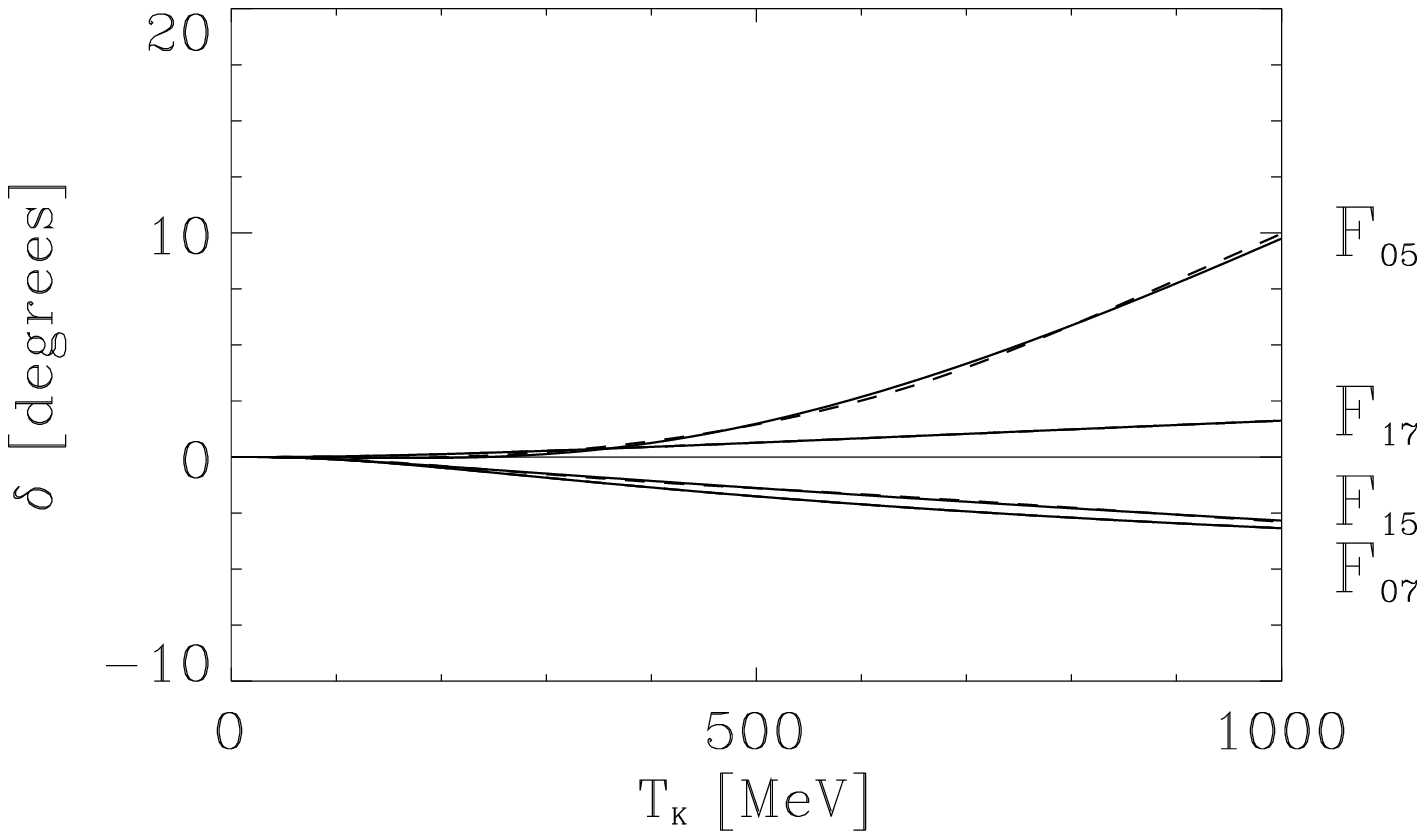,width=8cm}
\end{picture}
\begin{picture}(8,8)(0.0,0.0)\centering
\epsfig{figure=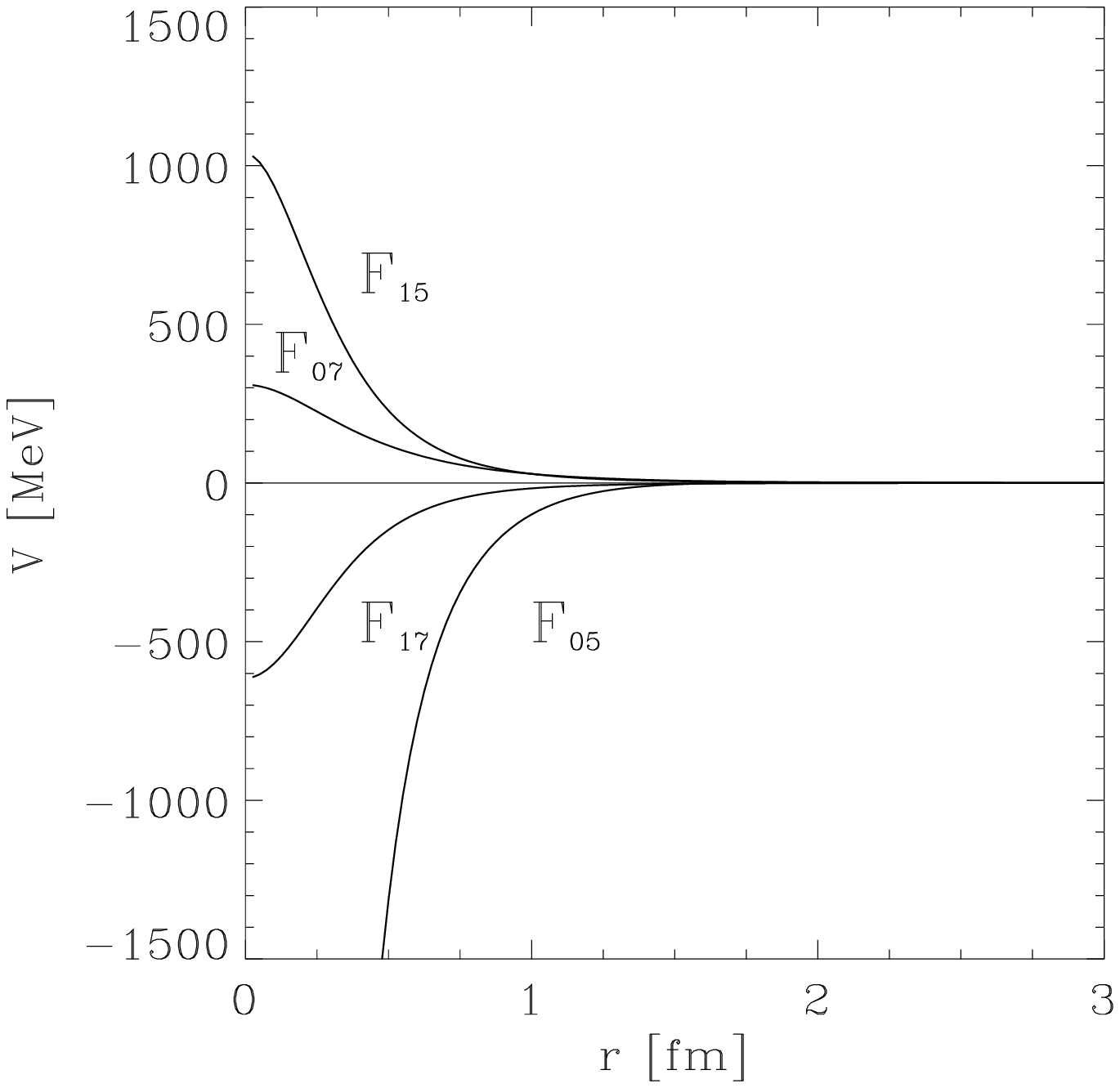,width=8cm}
\end{picture}
\caption%
{Left: $K^+N$ $\ell=3$ phase shifts from the analysis SP92 \protect\cite{arnkn}
(dashed) and their reproduction by the respective inversion potentials (full
line).
Right:
$K^+N$ $\ell=3$ inversion potentials.}
\label{abb_kn1}
\end{figure}

\unitlength 1cm
\begin{figure}[ht]\centering
\begin{picture}(8,8)(0.0,0.0)\centering
\epsfig{figure=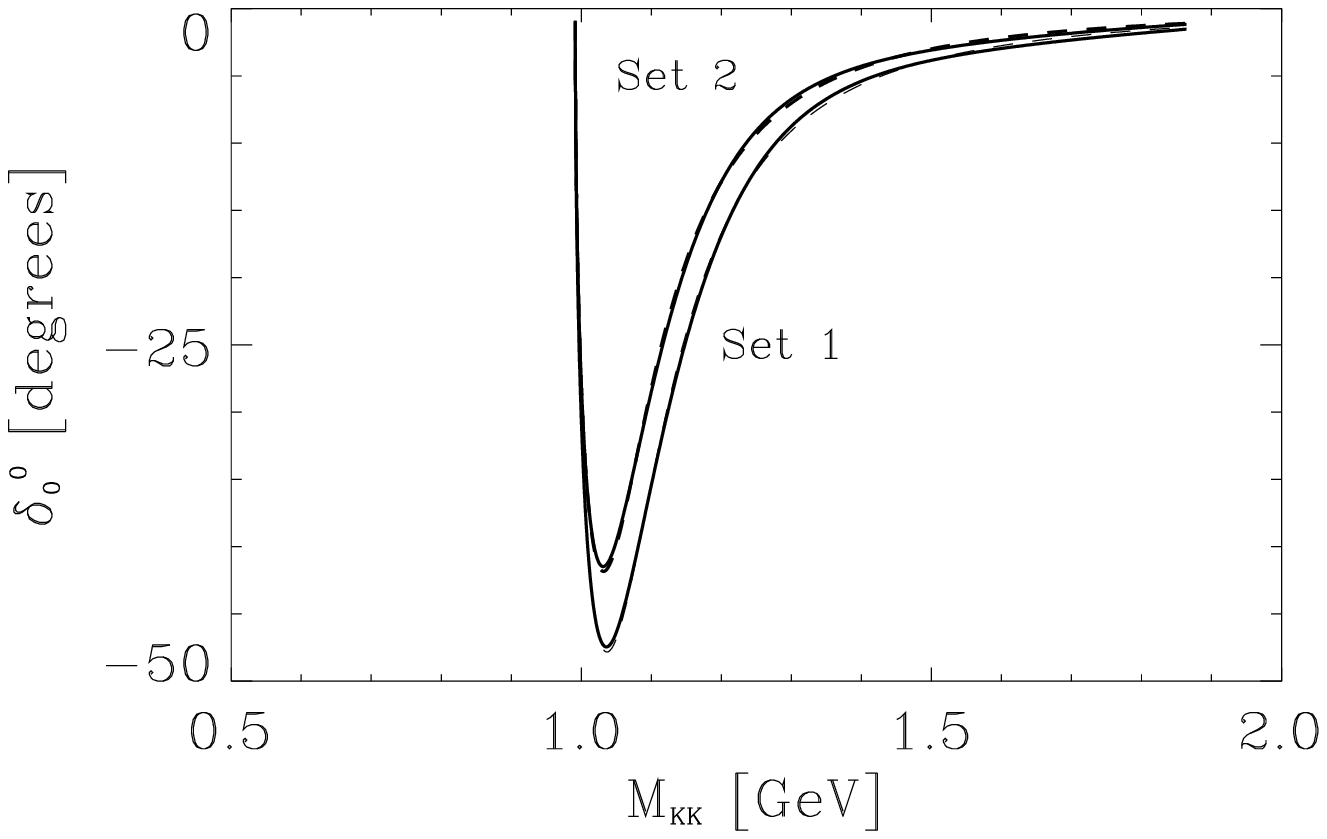,width=8cm}
\end{picture}
\begin{picture}(8,8)(0.0,0.0)\centering
\epsfig{figure=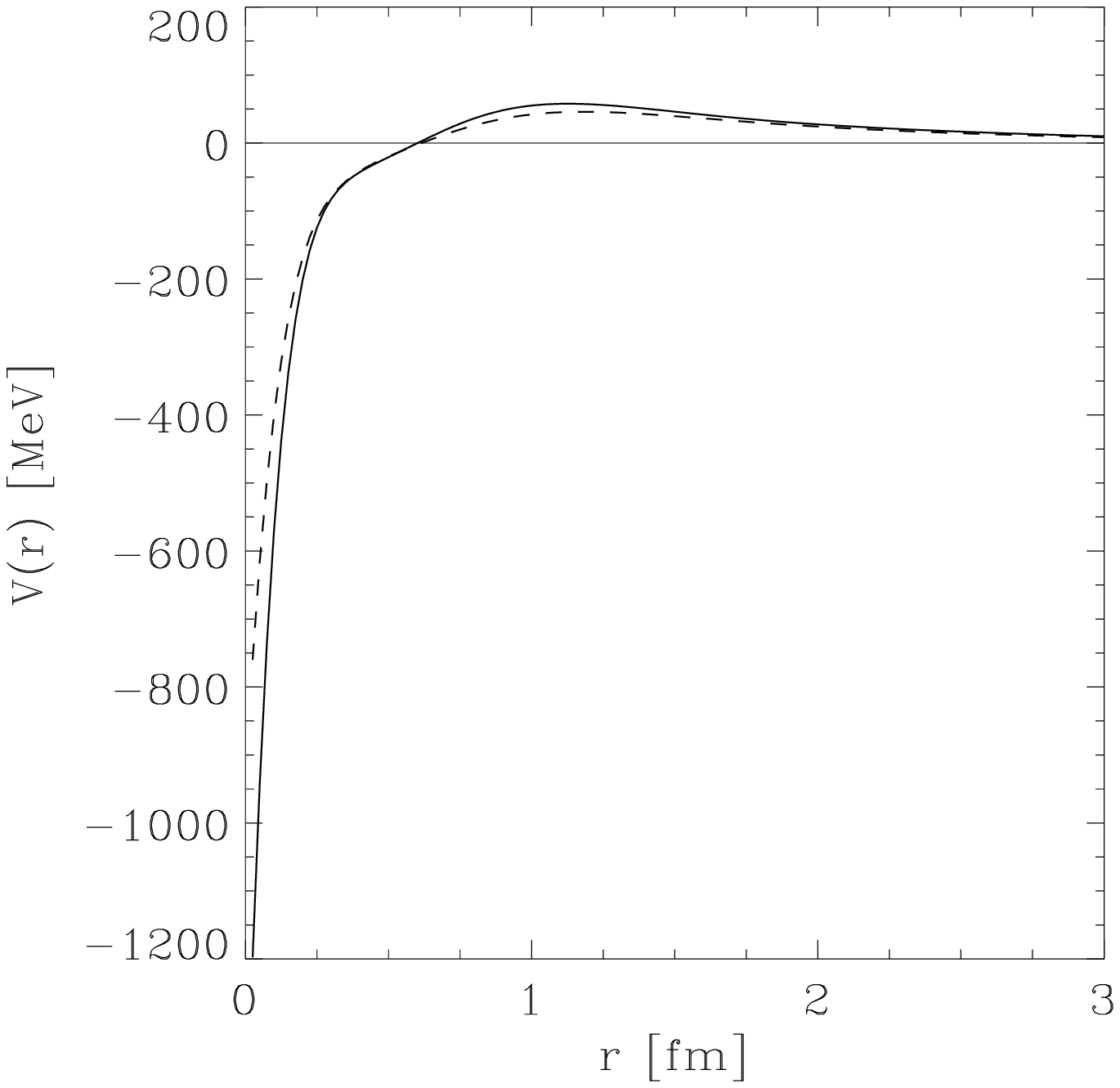,width=8cm}
\end{picture}
\caption%
{Left: $K\bar{K}$ $\ell=0$ real isoscalar phase shifts calculated
from the effective range expansion (dashed) using the
parameters given by Kaminski and Lesniak
\protect\cite{kam94} and their
reproduction by the inversion potentials (full line).
Right: $K\bar{K}$ $\ell=0$ real isoscalar potentials from quantum inversion
based on the two sets of parameters given by Kaminski and Lesniak
\protect\cite{kam94}, set 1: full line, set 2: dashed.}
\label{abb_kkbar1}
\end{figure}


\begin{references}
%
%
\bibitem{jae96}
L. J\"ade, M. Sander and H.V. von Geramb, {\it these conference proceedings} 
and nucl--th/9609054

\bibitem{cha89}
K. Chadan and P.C. Sabatier, {\it Inverse Problems in Quantum Scattering
Theory, 2nd. Edition}, Springer, Berlin (1989)

\bibitem{invnum}
H.V. von Geramb and H. Kohlhoff, in: H.V. von Geramb (ed.),
{\it Quantum Inversion Theory and Applications}, Lecture Notes in Physics
427, Springer, Berlin (1994); H. Kohlhoff and H.V. von Geramb, {\it ibid.}

\bibitem{san96}
M. Sander, PhD Thesis (in german), University of Hamburg (1996) 

\bibitem{arnpin}
R.A. Arndt, I.I. Strakowsky, R.L. Workman and M.M. Pavan,
Phys. Rev. {\bf C52}, 2120 (1995)

\bibitem{arnkn}
J. S. Hyslop, R. A. Arndt, L. D. Roper and R. L. Workman, Phys. Rev. {\bf
D46}, 961 (1992)

\bibitem{fro77}
C.D. Froggatt and J. L. Petersen, Nucl. Phys. {\bf B129}, 89 (1977)

\bibitem{est78}
P. Estabrooks et al., Nucl. Phys. {\bf B133}, 490 (1978)

\bibitem{kam94}
R. Kaminski and L. Lesniak, Phys. Rev. {\bf C51}, 2264 (1995)

\bibitem{bak86}
B.L.G. Bakker and P.J. Mulders, Adv. Nucl. Phys. {\bf 17}, 1 (1986)

\bibitem{kuk89}
V.I. Kukulin, V.M. Krasnopol'sky und J. Hor\'{a}\v{c}ek,
{\it Theory of Resonances}, Academia, Prague (1989)

\bibitem{koch80}
R. Koch and E. Pietarinen, Nucl. Phys. {\bf A336}, 331 (1980)

\bibitem{sigg95}
D. Sigg et al., Phys. Rev. Lett. {\bf 75}, 3245 (1995)

\bibitem{pea91}
B.C. Pearce and B.K. Jennings, Nucl. Phys. {\bf A528}, 655 (1991)

\bibitem{sch94a}
C. Sch\"utz, J.W. Durso, K. Holinde and J. Speth,
Phys. Rev. {\bf C49}, 2671 (1994)

\bibitem{GMO}
M.L. Goldberger, H. Miyazawa and R. Oehme, Phys. Rev. {\bf 99}, 986 (1955)

\bibitem{wor92}
R.L. Workman, R.A. Arndt and M.M. Pavan,  Phys. Rev. Lett. {\bf 68}, 1653
(1992)

\bibitem{loh90}
D. Lohse, J.W. Durso, K. Holinde and J. Speth,
Nucl. Phys. {\bf A516}, 513 (1990)

\bibitem{gas83}
J. Gasser and H. Leutwyler, Phys. Lett. {\bf 125B}, 325 (1983)

\bibitem{afa93}
L.G. Afanasyev et al., Phys. Lett. {\bf B308}, 200 (1993);
L.G. Afanasyev et al., Phys. Lett. {\bf B338}, 478 (1994)

\bibitem{efimov}
G.V. Efimov, M.A. Ivanov and V.E. Lyubovitskii,
Sov. J. Nucl. Phys. {\bf 44}, 296 (1986)

\bibitem{pionium} 
M. Sander, C. Kuhrts and H.V. von Geramb, Phys. Rev. {\bf C53}, R2610 (1996)

\bibitem{toer96}
N.A. T\"ornqvist and M. Roos, Phys. Rev. Lett. {\bf 76}, 1575 (1996)

\bibitem{isg96}
N. Isgur and J. Speth, Phys. Rev. Lett. {\bf 77}, 2332 (1996);
N.A. T\"ornqvist and M. Roos, {\em ibid.}, 2333 (1996)

\bibitem{hara96}
M. Harada, F. Sannino and J. Schechter, Los Alamos e--print archive
hep--ph/9609428; N.A. T\"ornqvist and M. Roos, {\em ibid.} hep--ph/9610527

\bibitem{ishi96}
S. Ishida et al.,
Los Alamos e--print archive hep--ph/9610359

\bibitem{bue90}
R. B\"uttgen, K. Holinde, A. M\"uller--Groeling, J. Speth and
P. Wyborny, Nucl. Phys. {\bf A506}, 586 (1990)

\bibitem{ber91a}
V. Bernard, N. Kaiser and U.--G. Meissner,
Phys. Rev. {\bf D43}, R2757 (1991)

\bibitem{bar92}
T. Barnes, E.S. Swanson and J. Weinstein,
Phys. Rev. {\bf D46}, 4868 (1992)

\bibitem{li94}
Z. Li, M. Guidry, T. Barnes and E.S. Swanson,
Bulletin Board hep--ph/9401326

\bibitem{kar80}
A. Karabarbounis and G. Shaw,
J. Phys. {\bf G6}, 583 (1980)

\bibitem{fie95}
H. R. Fiebig, H. Markum, A. Mih\'aly, K. Rabitsch and C. Starkjohann,
Nucl. Phys. B (Proc. Suppl.) {\bf 47}, 394 (1996)

\end{references}
\end{document}